\documentclass[twocolumn,showpacs,preprintnumbers,amsmath,amssymb,epsfig,widetext]{revtex4}

\usepackage{graphicx}
\usepackage{dcolumn}
\usepackage{bm}
\usepackage{epsfig}

\input epsf

\def\be{\begin{equation}}
\def\ee{\end{equation}}
\def\ba{\begin{eqnarray}}
\def\ea{\end{eqnarray}}

\newcommand{\potential}{{A}}
\newcommand{\shift}{{B}}
\newcommand{\curvature}{{H}_L}
\newcommand{\shear}{{H}_T}
\newcommand{\kh}{k_H}
\newcommand{\ck}{c_K}
\newcommand{\tppi}{\ck p_T \Pi_T}

\begin{document}

\preprint{}

\title{Parameterized Post-Friedmann Signatures of Acceleration in the CMB}

\author{Wayne Hu}
\email{whu@background.uchicago.edu}
\affiliation{Kavli Institute for Cosmological Physics, Department of Astronomy, \& Astrophysics, Enrico Fermi
Institute,  University of Chicago, Chicago IL 60637
}

\date{\today}

\begin{abstract}
We extend the covariant, parametrized post-Friedmann (PPF) treatment of cosmic acceleration
from modified gravity 
 to an arbitrary admixture of
matter, radiation, relativistic components and spatial curvature.  This generalization facilitates
the adaptation of Einstein-Boltzmann codes for solving CMB and matter perturbations in the linear
regime.   We use such a code to study the effect of metric evolution on the CMB through the
integrated Sachs-Wolfe effect.   We discuss the ability of modified gravity to alter the low multipole
spectrum, including lowering the power in the quadrupole.
 From a principal component description of the primary metric
ratio parameter, we obtain general constraints from WMAP on modified gravity models of the
acceleration. 
\end{abstract}


\maketitle

\section{Introduction}
\label{sec:introduction}

In the absence of theoretically compelling models for the acceleration of the expansion,
it is useful to have a parametrized description of possible deviations from the standard
cosmological constant model.   To explore modified gravity explanations of the acceleration
we need a parametrized post-Friedmann (PPF) description of gravity that parallels the 
parametrized post-Newtonian description of solar system tests.

Many attempts to parametrize deviations phenomenologically 
 (e.g.~\cite{IshUpaSpe06,KnoSonTys06,WanHuiMayHai07,Son06,HutLin07}) 
 or across a limited range of scales
 (e.g.~\cite{CalCooMel07,Amendola:2007rr,ZhaLigBeaDod07,Jain:2007yk,AmiWagBla07}) exist in the
 literature.  Hu \& Sawicki \cite{HuSaw07b} recently introduced a PPF approach
 that spans all linear scales and includes an ansatz for non-linear phenomenology.  
 The linear framework is based on enforcing the requirements of a 
 metric theory, with small deviations from the Friedmann-Robertson-Walker metric,
 and covariant conservation of the stress energy tensor.
  This approach provides an
 excellent description of at least two specific models of modified gravity, the self-accelerating
 branch of the
 braneworld acceleration model (DGP \cite{DvaGabPor00}) and the modified action
 $f(R)$ model \cite{Caretal03,NojOdi03,Capozziello:2003tk}.
 
In this Paper, we extend the PPF approach to include relativistic matter components
and spatial curvature.  This generalization yields a self-consistent means of evolving metric
perturbations from initial conditions in the radiation dominated era through to the present.
We also phrase the PPF description in a manner that is useful for adapting
an Einstein-Boltzmann code (e.g.~\cite{SelZal96,Lewetal00}) for modified gravity.
Furthermore,  specific modified gravity models already require such components at low
redshift,
e.g. the self-accelerating branch of DGP requires spatial curvature to fit
distance measures \cite{SonSawHu06} and the normal branch requires dark energy
to explain acceleration.

We use this generalization to study the impact of evolution in the metric on the
cosmic microwave background (CMB) through the integrated Sachs-Wolfe (ISW)
effect.  In the DGP and $f(R)$ models of cosmic acceleration it is well known
that the ISW effect provides one of the most powerful
and robust tests available  \cite{LueScoSta04,SonSawHu06,SonPeiHu07,Schmidt:2007vj,CalCooMel07,Pogosian:2007sw}.  Since it arises from fluctuations near the horizon scale
during the acceleration epoch, it  requires a fully covariant description such as the
one developed here.    It also therefore tests gravity on the largest scales observable.

We begin in \S \ref{sec:formalism} with the multicomponent generalization of the PPF
description.  This derivation is given in the comoving gauge but we describe in the Appendix
the corresponding relations in the synchronous gauge.
In \S \ref{sec:ISW} we explore the phenomenology of the ISW effect in the
CMB and examine current constraints.   We discuss these results in \S \ref{sec:discussion}.
 
\section{Multicomponent Formalism}
\label{sec:formalism}
 
In \S \ref{sec:covariant}, we adapt the structure of multicomponent covariant linear perturbation theory 
to modified gravity models.  This treatment reduces to that of \cite{HuSaw07b} in the limit of
a flat, matter only universe.  We discuss gauges relevant for the PPF parametrization in
\S \ref{sec:gauge} and derive the explicit representation in \S \ref{sec:ppf}.

\subsection{Covariant Structure}
\label{sec:covariant}

For the purposes of constructing a self-consistent covariant parametrization of
modified gravity, it is convenient to view the modifications to gravity
in terms of an additional ``dark energy" stress  tensor \cite{Kunz:2006ca,CalCooMel07}.  Given a metric theory of
gravity,
we are always free to {\it define} the dark energy stress tensor to be 
\begin{equation}
T^{\mu\nu}_{e} \equiv {1 \over 8\pi G} G^{\mu\nu} - T^{\mu\nu}_{T} \,,
\label{eqn:effectivede}
\end{equation}
where we use the subscript ``$T$" for the total stress energy tensor, which combines all true
components, and ``$e$" 
for the effective dark energy here and
below.  Note that we allow the total stress energy tensor to contain true dark
energy components such as a cosmological constant or scalar fields
 for maximal generality
({\it cf.} \cite{HuSaw07b}).

The effective dark energy behaves as a separately conserved system by virtue of
the Bianchi identities and the conservation of $T^{\mu\nu}_{T}$
\cite{Bashinsky:2007yc}
\begin{equation}
\nabla_\mu T_e^{\mu\nu} = {1 \over 8\pi G} \nabla_\mu G^{\mu\nu} - \nabla_\mu T^{\mu\nu}_{T}  =0 \,.
\end{equation}
However the closure relation for this conservation law, which specifies the
relationship between the components of the stress energy tensor, can and will in general 
depend on the matter content in contrast to minimally coupled dark energy models
\cite{Hu98,HuSaw07b,Bashinsky:2007yc}.

As usual, statistical isotropy requires that at the background level the stress tensors
can be parametrized by the energy density and pressure
\begin{align}
{T^0_{\hphantom{0}0}} &= -\rho\, , \nonumber\\
{T_0^{\hphantom{i}i}} &= 0 \,, \nonumber\\
{T^i_{\hphantom{i}j}} &= {p}\delta^i_{\hphantom{i}j} \,, 
\label{eqn:stressenergy}
\end{align}
for both the total and the effective stress energy tensor. 

The background Einstein tensor is built out of the Friedmann-Robertson-Walker metric
\begin{align}
ds^{2} &= g_{\mu\nu} dx^{\mu}dx^{\nu}= a^{2} ( -d\eta^{2} + \gamma_{ij}dx^{i} dx^{j} )\,, 
\end{align}
where $\eta = \int dt/a$ is the conformal time. In spherical coordinates $\gamma_{ij}$,
the spatial metric with constant curvature $K$,  
can be represented as
\begin{align}
\gamma_{ij}dx^{i}dx^{j} &= d D^{2} + D_{A}^{2}d\Omega\,,
\end{align}
where 
$D_{A}=K^{-1/2} \sin( K^{1/2} D )$ is the angular diameter distance. 
The Einstein equation (\ref{eqn:effectivede}) becomes the usual Friedmann equation 
\begin{equation}
H^2 + {K \over a^{2}} = {8\pi G \over 3} (\rho_T+\rho_e)\,.
\end{equation}
The conservation laws 
\begin{align}
\rho_T' &= -3(\rho_T+p_T) \,, \nonumber\\
\rho_e' &= -3(\rho_e+p_e)\,,
\end{align}
close the system of equations for the background equations.  The effective dark energy is
thus parametrized in the same way as true dark energy: by the density today in units
of critical $\Omega_{e} = 8\pi G \rho_{e}(\ln a=0)/3 H_{0}^{2}$ and the effective equation of state
$w_{e}(\ln a) = p_{e}/\rho_{e}$.

Scalar linear perturbations may be decomposed into the
eigenfunctions of the Laplace operator
\begin{align}
\nabla^2 Y = -k^2 Y \,,
\end{align}
and its covariant derivatives 
\begin{align}
Y_i &  = (-k) \nabla_i Y \,,\nonumber\\ 
Y_{ij} &= (k^{-2} \nabla_i \nabla_j + \gamma_{ij}/3) Y \,.
\end{align}
The most general scalar linear perturbations to the metric of a wavenumber $k$
can be parametrized
by \cite{Bar80,KodSas84}
\begin{align}
\delta {g_{00}} &= -a^{2} (2 {\potential}Y)\,, \nonumber\\
\delta {g_{0i}} &= -a^{2} {\shift} Y_i\,,  \nonumber\\
\delta {g_{ij}} &= a^{2} (
        2 {\curvature} Y \gamma_{ij} + 2 {\shear Y_{ij}})\,.
\label{eqn:metric}
\end{align}
Likewise the stress energy tensors can be parametrized as
\begin{align}
\delta{T^0_{\hphantom{0}0}} &=  - { \delta\rho}\,, \nonumber\\
\delta{T_0^{\hphantom{i}i}} &= -(\rho + p){v}Y^i\,, \nonumber\\
\delta {T^i_{\hphantom{i}j}} &= {\delta p}Y  \delta^i_{\hphantom{i}j} 
	+ p{\Pi Y^i_{\hphantom{i}j}}\,.
\label{eqn:dstressenergy}
\end{align}
We again describe the modification to gravity with an effective dark energy
stress tensor and allow the total stress energy tensor to be composed of 
multiple components
\begin{align}
\delta\rho_T &= \sum_i \delta \rho_i \,, \nonumber\\
(\rho_T+p_T) v_T  &= \sum_i (\rho_i+p_i) v_i \,,  \nonumber\\
\delta p_T &= \sum_i \delta p_i \,,\nonumber\\
p_T\Pi_T &= \sum_i p_i\Pi_i \,.
\end{align}
By definition, Eqn.~(\ref{eqn:effectivede}) enforces the usual 4 Einstein
field equations \cite{HuEis99}
\begin{align}
& {H_L}+ {1 \over 3} {H_T}+   {B \over \kh}-  {H_T' \over \kh^2} 
   \nonumber\\&\qquad
   = { 4\pi G \over H^2 \ck \kh^{2}}   \left[ {\delta \rho} + 3  (\rho+p){{v}-
{B} \over \kh }\right] \,,
\nonumber\\
&  {A} + {H_L} +  {H_T \over 3} + 
 {B'+2B \over \kh} - 
 \left[ {H_T'' \over \kh^2} + \left( 3 + {H'\over H}\right) {H_T' \over \kh^2} \right]
   \nonumber\\&\qquad
 = -{8\pi G \over H^2 \kh^2} {p\Pi}  \,,
\nonumber\\
& {A} - {H_L'} 
- { H_T' \over 3} - {K \over (aH)^2} \left( {B \over k_H} - {H_T' \over k_H^2} \right)
   \nonumber\\&\qquad
   =  {4\pi G \over H^2 } (\rho+p){{v}-{B} \over \kh} \,,
\nonumber\\
 & A' + \left( 2 + 2{H' \over H} - {\kh^2 \over 3}  \right) A 
-{\kh \over 3}  (B'+B) \nonumber\\&\qquad - H_L'' - \left( 2 + {H' \over H}\right) H_L' 
   = {4\pi G \over H^2} ({\delta p} + {1 \over 3}{\delta\rho} ) \,,
\label{eqn:einstein}
\end{align}
where the stress energy components on the rhs combine the total and effective 
contributions.
Here $'=d/d\ln a$, $\kh = (k/aH)$, and $c_K = 1-3K/k^2$.   The conservation laws
 become the continuity and Navier-Stokes equations
\begin{align}  \label{eqn:conservation}
& {\delta\rho_i'}
	+  3({\delta \rho_i}+ {\delta p_i})
=
        -(\rho_i+p_i)(\kh {v}_i + 3 H_L')\,, \\
&      {[a^4(\rho_i + p_i)({{v_i}-{B}})]' \over a^4\kh}
= 
 { \delta p_i }- {2 \over 3}\ck p_i {\Pi_i} + (\rho_i+ p_i) {A} \,, \nonumber
\end{align}
for all components separately conserved including the effective dark energy.

\subsection{Gauge}
\label{sec:gauge}

The covariant Einstein and conservation equations (\ref{eqn:einstein}) and
(\ref{eqn:conservation}) apply to any choice of gauge.  For the PPF construction
it is useful to work with variables that take on certain meanings such as curvature
and potential fluctuations in specific gauges.  For numerical codes it is useful to
have a covariant representation of such variables so that they may be accessed
from other gauge choices.

 Under
a gauge transformation defined by the change in conformal time slicing $T$
and spatial threading $L$ \cite{Bar80,KodSas84}
\begin{align}
 \eta &=\tilde \eta +{T}\,, \\
x^i   &= \tilde x^i +{L}Y^i\,, \nonumber
\end{align}
the metric  variables transform as
\begin{align}
 A &=\tilde A - aH (T'+T)  \,, \nonumber\\
 B &=  \tilde B +  aH( L' + k_H {T}) \,, \nonumber\\
 H_L &=  \tilde H_L - aH(T + {1\over 3}k_H L) \,, \nonumber\\
 H_T &=  \tilde H_T + aH k_H  {L}\,, 
\label{eqn:metrictrans}
\end{align}
and the stress energy components transform as
\begin{align} 
{\delta \rho} &= \widetilde{\delta\rho} - \rho' aH {T}\,, \nonumber\\ 
{\delta  p} &= \widetilde{\delta p} -  p' aH  {T}\,, \nonumber\\
 v &=  \tilde v +   aH { L'} \,, \nonumber\\
 \Pi &= \tilde \Pi \,.
\label{eqn:fluidtrans}
\end{align}
A gauge is fully specified if the functions $T$ and $L$ are uniquely defined.

Following \cite{HuSaw07b}, we shall construct the PPF description from a combination
of (total matter or) comoving and Newtonian gauge quantities.
The comoving gauge is specified by the conditions
\begin{align}
B &= v_T \,, \nonumber\\
H_T &= 0 \,.
\end{align} 
They fully specify the gauge transformation from an alternate gauge
choice
\begin{align}
T &= (\tilde v_T - \tilde B)/k  \,, \nonumber\\
L &=  -\tilde H_T /k \,.
\end{align}
To avoid confusion between fluctuations defined in different gauges, we will assign special variable
names to comoving gauge quantities
\begin{eqnarray}
\zeta &\equiv& H_L \,,\nonumber\\
\xi &\equiv& A  \,,\nonumber\\
 \rho\Delta  &\equiv& \delta \rho \,, \nonumber\\
\Delta p &\equiv& \delta p \,, \nonumber\\
V &\equiv& v \,.
\end{eqnarray}
$\Delta p$ should not be confused with $p \Delta = p (\delta \rho/\rho)$.

Similarly, the Newtonian
gauge is defined by the condition $B=H_T=0$ and the transformation 
\begin{align}
T &= -{\tilde B \over k} + { \tilde H_T'  \over k \kh } \,, \nonumber\\
L &=  -{\tilde H_T \over k} \,.
\end{align}
To avoid confusion we define
\begin{align}
\Phi &\equiv H_L \,,\nonumber\\
\Psi &\equiv A \,.
\end{align}
The relationships between the two metric fluctuations are 
\begin{align} \label{eqn:comnewt1}
\zeta &= \Phi - {V_T \over \kh} \,, \\
\xi & = \Psi - { V_T' + V_T \over \kh} \,.
\label{eqn:comnewt2}
\end{align}
We refrain from utilizing  matter variables in Newtonian gauge but note that
velocities in the two gauges are the same.    We discuss the synchronous gauge
representation in the Appendix.

\subsection{PPF Parameterization}
\label{sec:ppf}
With the full covariant framework of linear perturbation theory in place,
we now generalize the PPF description of the effective dark energy stress
tensor \cite{HuSaw07b} for multiple relativistic components and spatial curvature.

Following the original construction, we demand that the additional PPF contribution 
satisfy two requirements. 
On superhorizon scales $k_H \equiv k/aH \ll 1$, the curvature $\zeta$ in the
comoving gauge is only altered by the effective dark energy at second order in $k_H$.
Conservation of $\zeta$ in the absence of curvature and non-adiabatic stress fluctuations
is a consequence of energy-momentum conservation 
\cite{HuEis99} and applies to modified gravity
models that satisfy it \cite{Ber06}.

The third Einstein equation (\ref{eqn:einstein}) reads 
\begin{align}
\zeta' = \xi - {K \over (aH)^2}{V_T \over k_H}
- {4\pi G \over H^2} (\rho_e + p_e) {V_e - V_T \over k_H}\,,
\label{eqn:zetaprimegeneral}
\end{align}
and the Navier-Stokes equation for the total velocity (\ref{eqn:conservation}) gives
\begin{align}
\xi =  -{\Delta p_T - {2\over 3}\tppi \over \rho_T + p_T}  \,.
\label{eqn:xieom}
\end{align}
Without further loss of generality we can parametrize the effective dark energy contribution in this limit by a function $f_\zeta(a)$
where
\begin{equation}
\lim_{k_H \ll 1}
 {4\pi G \over H^2} (\rho_e + p_e) {V_e - V_T \over k_H}
= - {1 \over 3} \ck  f_\zeta(a) k_H V_T
\end{equation}
since $V_T = {\cal O}(k_H \zeta)$ for adiabatic fluctuations.
The equation of motion for $\zeta$ in this limit now reads
\begin{align}
\lim_{k_H \ll 1} \zeta' & = -{\Delta p_T - {2\over 3}\tppi \over \rho_T + p_T} - {K \over k^2} k_H V_T \nonumber\\
& \quad +{1 \over 3} \ck  f_\zeta k_H V_T \,.
\label{eqn:zetaprimesh}
\end{align}

The second condition is that the metric satisfies a Poisson-like equation in the 
$k_H \gg 1$ quasistatic limit
\begin{equation}
\lim_{k_{H}\gg 1} \Phi_- = {4\pi G \over \ck k_H^2 H^2}{\Delta_T \rho_T + \tppi\over  1+f_G(a)} \,,
 \label{eqn:qspoisson}
\end{equation}
where $f_G$ depends on time alone and $\Phi_- \equiv (\Phi - \Psi)/2$.
To make these two limits compatible we introduce a parameter $\Gamma$ such that
\begin{equation}
\Phi_-+\Gamma = {4\pi G
\over \ck k_H^2 H^2} [\Delta_T \rho_T + \tppi]
\label{eqn:modpoiss}
\end{equation}
on all (linear) scales. 
Comparison with the first Einstein equation (\ref{eqn:einstein}) for the Newtonian metric
perturbations gives the first closure relation for the effective dark energy
\begin{equation}
\rho_{e}\Delta_{e} + 3(\rho_{e}+p_{e}) {V_{e}-V_{T}\over k_{H} } + \ck p_{e}\Pi_{e} = 
-{k^{2}\ck \over 4\pi G a^{2}} \Gamma \,.
\label{eqn:ppffluid}
\end{equation}
Given that modified gravity models are most simply parametrized in terms of the
relationship between $\Phi$ and $\Psi$ that they induce, we describe 
the second closure condition through the 
effective anisotropic stress 
\begin{equation}
\Phi_+ \equiv {\Phi + \Psi \over 2} = g(a,k) \Phi_- - {4\pi G \over H^2 k_H^2} p_{T}\Pi_{T} \,.
\label{eqn:gdef}
\end{equation}
Parametrizing this relation with a  free function $g(a,k)$ again loses no further
generality ({\it cf}. \cite{CalCooMel07}).  Without a modification to
gravity $g=0$.  If the true anisotropic stress is negligible then $g = \Phi_+/\Phi_-$ and for
this reason we will refer to it as the metric ratio parameter.  
Equation~(\ref{eqn:gdef}) defines the effective anisotropic stress as
\begin{equation}
{4 \pi G \over H^2 k_H^2}p_e \Pi_e = -g \Phi_- \,.
\label{eqn:pieff}
\end{equation}

To complete these equations we must determine an equation of motion for $\Gamma$
that is consistent with the two requirements (\ref{eqn:zetaprimesh}) and 
(\ref{eqn:qspoisson}).
Taking the derivatives of Eqn.~(\ref{eqn:comnewt1}) and (\ref{eqn:modpoiss}) with the
help of the continuity and Navier-Stokes equations,
we obtain the condition at $k_H \ll 1$
\begin{eqnarray}
\lim_{k_H \ll 1} \Gamma'  = S -\Gamma \,,
\end{eqnarray}
where the source
\begin{align}
S&={g' -2 g \over g+1} \Phi_- +  {4\pi G \over (g+1)k_H^2 H^2} \Big\{ g [(p_T\pi_T)' + p_T\pi_T]\nonumber\\
&\quad - 
 \left[ (g + f_\zeta + g f_\zeta)(\rho_T+p_T) - (\rho_e+p_e) \right] k_H V_T \Big\}\,.
\end{align}
Note that $S=0$ if the modification to gravity vanishes: $g=0$ and $f_\zeta=0$.

In the opposite limit a comparison of Eqns.~(\ref{eqn:qspoisson}) and (\ref{eqn:modpoiss}) imply
\begin{equation}
\lim_{k_H \gg 1} \Gamma = f_G \Phi_- \,.
\end{equation}
To satisfy both limits, we take the equation of motion for $\Gamma$ to be
\begin{equation}
(1 + c_\Gamma^2 k_H^2) [\Gamma' + \Gamma + c_\Gamma^2 k_H^2 (\Gamma - f_G \Phi_-)] = S\,.
\label{eqn:gammaeom}
\end{equation}
Substituting this relation into the derivative of the Poisson equation, we obtain
\begin{eqnarray}
\label{eqn:zetaeom}
\zeta' & = & -{\Delta p_T - {2\over 3}\tppi \over \rho_T + p_T} - {K \over k^2} k_H V_T \\
&&+ {g+1 \over F} \left[  S -\Gamma'-\Gamma + f_\zeta { 4\pi G (\rho_T + p_T) \over H^2} {V_T \over k_H} \right]  \nonumber
\end{eqnarray}
where 
\begin{equation}
F(a) = 1 + 3(g+1) {4 \pi G a^2 \over k^2 \ck} (\rho_T + p_T) \,.
\end{equation}
Comparison with Eqn.~(\ref{eqn:zetaprimegeneral}) implicitly
defines the effective dark energy momentum density 
\begin{align}
{ V_{e}-V_{T}\over k_{H}} &=-{H^{2} \over 4\pi G (\rho_{e} + p_{e}) } {g+1 \over F}
\label{eqn:veeff} \\
&\quad \times
\left[ S - \Gamma' - \Gamma + f_{\zeta}{4\pi G (\rho_{T}+p_{T}) \over H^{2}}{V_{T}\over k_{H}}
\right]  \,.\nonumber
\end{align}
Likewise the effective pressure can be defined from this relation through
the dark energy Navier-Stokes equation
\begin{equation}
{[a^{4}(\rho_{e}+p_{e})(V_{e}-V_T)]' \over a^{4}k_{H}} = \Delta p_{e}- {2\over 3}\ck p_{e}\Pi_{e} + (\rho_{e}+p_{e}) \xi \,.
\label{eqn:pi}
\end{equation}
This completes the multicomponent generalization of the PPF parametrization.  It is described by three free functions $f_\zeta(\ln a)$, $f_G(\ln a)$ 
and $g(\ln a,k)$ and one parameter $c_\Gamma$
in addition to the usual $w_e(a)= p_e(a)/\rho_e(a)$ which determines the background 
expansion.   In the limit that the universe is spatially flat and relativistic matter components
are subdominant, this description exactly matches the original treatment of \cite{HuSaw07b}.

In our numerical implementation, we evolve the density $\Delta_i$ and relative
velocity $V_i-V_T$ components through the comoving gauge conservation equations and the 
Boltzmann hierarchy for the radiation closure condition and anisotropy using
the code of \cite{HuSelWhiZal98,HuOka03}.
Specifically we evolve the comoving curvature fluctuation $\zeta$ through
Eqn.~(\ref{eqn:zetaeom}) and $\xi$ from the constraint Eqn.~(\ref{eqn:xieom}).
The final metric perturbation is the
 total velocity $V_T(=B)$ which is evolved using the combined Newtonian and comoving
Navier-Stokes equations  [see also the gauge relation (\ref{eqn:comnewt1})]
\begin{equation}
V_T' + V_T = k_H (\Psi-\xi)\,,
\end{equation}
where $\Phi$ and $\Psi$ are specified by Eqn.~(\ref{eqn:modpoiss}) and $g(\ln a,k)$.
Finally  $\Gamma$ is obtained by evolving Eqn.~(\ref{eqn:gammaeom}).  
The second gauge relation of Eqn.~(\ref{eqn:comnewt2}) between $\zeta$ and $\Phi$ is
used as a test of numerical accuracy.

Note that the relationship between $\Gamma$ and $g(\ln a,k)$ and the effective
energy density, anisotropic stress, momentum density and pressure
fluctuations given by Eqns.~(\ref{eqn:ppffluid}), (\ref{eqn:pieff}), (\ref{eqn:veeff})
and (\ref{eqn:pi}) are only implicitly used.  Likewise while the effective stress energy
obeys the continuity and Navier-Stokes equations, they are not used in the numerical
scheme.  These relations are however useful for adapting the numerical scheme for the
synchronous and 
other gauges  (see Appendix).

\begin{figure}[tb]
\begin{center}
\includegraphics[width=3.4in]{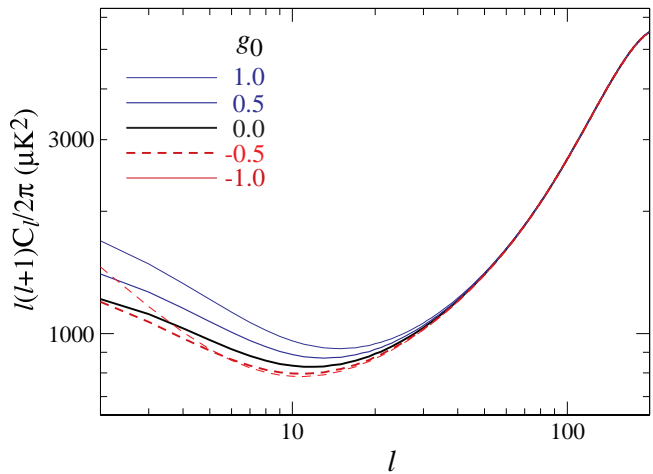}
\caption{CMB temperature power spectrum as a function of the amplitude of the
metric ratio parameter today $g_0$ given 
the evolution described by Eqn.~(\ref{eqn:gevolution}).  $\Lambda$CDM corresponds to $g_{0}=0$.  Increasing
$|g_{0}|$ in the positive direction monotonically increases the ISW effect at low multipoles 
whereas in the negative direction it first decreases and then increases the effect.
 The other parameters have been fixed at 
$c_{g}=0.01$, $c_{\Gamma}=1$ and $w_{e}=-1$.} \label{fig:g0}
\end{center}
\end{figure}

\section{PPF ISW Effect in the CMB}
\label{sec:ISW}

In general, our PPF description of cosmological deviations from general relativity includes  one 
function of space and time that interrelate the metric fluctuations 
$g(\ln a,k)$, two functions of time only $f_G(\ln a)$ and
$f_\zeta(\ln a)$ that determine the relationship between the metric and the matter,
 and one parametrized scale $c_\Gamma$ that bridges the transition between the latter
 two relations.  These free parameters allow a complex range of phenomena for the 
 scale-dependent evolution of linear perturbations.   The impact of such deviations 
on specific observables however limits the relevant parameters to a more manageable range.

Here we shall focus on the 
CMB temperature anisotropy induced by the evolution of the metric during the
acceleration epoch, the so-called the integrated Sachs-Wolfe (ISW) effect.   
This effect also highlights the extensions to the PPF description introduced in the previous
sections.  They allow an Einstein-Boltzmann code to self-consistently calculate the net CMB anisotropy from
the initial epoch where radiation dominates through to the present.    

Though the 
formalism allows for high redshift modifications of gravity, 
we work in the context that they only appear during the recent
acceleration epoch.  Hence the well-tested high multipole structure and polarization of the CMB
are left unchanged.  

In \S \ref{sec:phenomenology} we describe the impact of the PPF parameters on the low
multipole temperature anisotropy of the ISW effect.   We examine constraints on these
parameters from WMAP in \S \ref{sec:WMAP} and the reverse-engineering of models
to produce specific features in the spectrum in \S \ref{sec:designer}.

\subsection{Phenomenology}
\label{sec:phenomenology}

We begin by 
 specializing the PPF description for near horizon scale perturbations. 
The ISW effect is associated with the evolution of the metric potential $\Phi_{-}$ during the 
acceleration epoch of $z \lesssim 1$ and wavenumbers $k \sim 10^{-3}$ Mpc$^{-1}$.
Correspondingly, it is important to characterize the PPF  metric ratio parameter
$g(\ln a,k)$ at such epochs and scales.
Since $g(\ln a,k)$ 
should become independent of scale 
once $k_H \ll 1$ let us take the functional form
\begin{align}
 g(\ln a,k)  &= { g_{\rm SH}(\ln a) \over 1+ (c_{g}\kh)^{2}}\,.
 \label{eqn:ginterpolation}
\end{align}
 We shall use the parameter $c_g$ to explore the range of scales
that the ISW effect tests.  We take $g \rightarrow 0$ for $k_H \gg 1$ and $f_G=0$ so
as to restore general relativity on small scales.  Such models would evade all
current constraints from large-scale structure and are uniquely probed by the CMB.
On the other hand, we shall see that 
if $c_{g} \lesssim 0.01$ our constraints are valid for other choices 
since the high $k$ behavior of $g$ does not impact the ISW effect.  
In general, our treatment should be interpreted as constraining the average $g$ during the acceleration epoch and
for scales near the horizon regardless of its behavior on small scales.

Next, we assume that deviations in $g$ will only appear when the modification to 
gravity becomes an important contributor to the expansion rate $H \propto \rho^{1/2}$.
We therefore take a baseline functional form of
\begin{align}
g_{\rm SH}(\ln a) = g_0 \left( {\rho_e  \over \rho_T} {\Omega_T \over \Omega_e} \right)^{1/2}
\label{eqn:gevolution}
\end{align}
such that $g_{\rm SH}(0)=g_0$.  In the self-accelerating branch of DGP, $g$ grows to 
$g_0 \sim 1$ by the present on scales near the horizon and in the $f(R)$ models $g = -1/3$ on
scales below a Compton wavelength that increases with time \cite{HuSaw07b}. 
In the next section we will explore variations from the
baseline functional form of Eqn.~(\ref{eqn:gevolution}).  

\begin{figure}[tb]
\begin{center}
\includegraphics[width=3.4in]{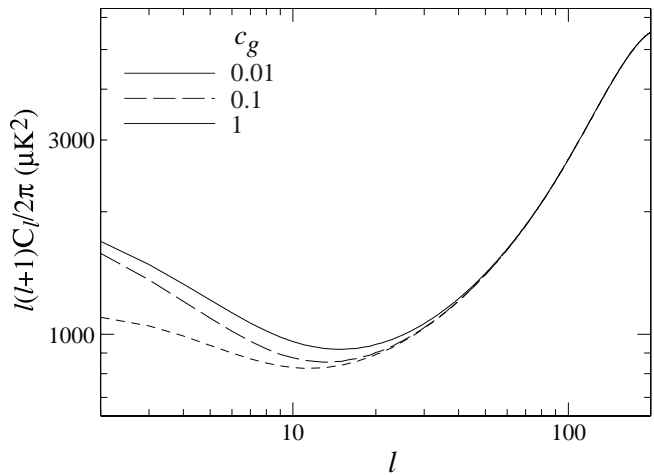}
\caption{ Effect of changing the scale at which metric ratio deviations occur through $c_g$.  Once
$c_{g}\gtrsim 0.1$ deviations have been suppressed for $k \gtrsim 10^{{-3}}$Mpc$^{-1}$ where the ISW
effect at the lowest multipoles peak. Other parameters have been fixed to $g_{0}=1$, $c_{\Gamma}=1$
and $w_{e}=-1$.} \label{fig:cg}
\end{center}
\end{figure}

The remaining important parameter is $c_\Gamma$ which defines in units of the Hubble scale
where the transition to the quasi-static Poisson equation  occurs.  We begin by 
choosing $c_\Gamma=1$ as motivated by the DGP and $f(R)$ models \cite{HuSaw07b}.

Finally for definiteness we take $f_\zeta(\ln a) = 0.4 g_{\rm SH}(\ln a)$ for the relationship
between the density $\Delta$ and metric fluctuations $\Phi_-$ on superhorizon scales
(see \cite{HuSaw07b} for a discussion).  
The ISW effect is insensitive to variations in this choice if $c_\Gamma \approx 1$.

We take $w_e=-1$ to illustrate effects that are coming purely from the modification
of gravity and not the change in the expansion history.  For the other cosmological
parameters, we take the maximum likelihood values for WMAP3 \cite{Speetal06}: 
$\Omega_{m}h^{2}=0.128$,
$\Omega_{b}h^{2}=0.0223$, $\Omega_{e}=0.76$, $\tau=0.092$ with an initial power spectrum
of 
\begin{equation}
{k^{3}P_{\zeta} \over 2\pi^{2}}=\delta_{\zeta}^{2}\left( {k \over 0.05{\rm Mpc}^{-1} }\right)^{n-1}
\label{eqn:pzeta}
\end{equation}
with $\delta_{\zeta}=4.56 \times 10^{-4}$ and $n=0.958$.

We begin by exploring the impact of the amplitude of the metric deviation $g_{0}$ in
Fig.~\ref{fig:g0} with the other parameters set to their fiducial values: 
$c_{g}=0.01$, $c_{\Gamma}=1$, $w_{e}=-1$.  
Note that $g_{0}=0$ is exactly equivalent to the $\Lambda$CDM model.
For $g_0 >0$ (solid lines), the
ISW contributions increase monotonically since raising $g_0$ corresponds to increasing the
 decay of the potential $\Phi_{-}$. For $g_0<0$, the decay is slowed and eventually turns into growth.
Correspondingly, the ISW effect is first reduced and then enhanced as $|g_0|$ increases.

In Fig.~\ref{fig:cg}, we illustrate the effect of suppresing $g$ at high $k$
through $c_g$ in Eqn.~(\ref{eqn:ginterpolation}) for $g_{0}=1$, $c_{\Gamma}=1$ and $w_{e}=-1$.  As $c_g$ is raised beyond $0.1$, the 
enhancement to the ISW effect begins to go away.  This confirms that the relevant
scales at which $g$ is constrained from the effect is near a tenth of the horizon scale
or $k \sim 10^{-3}$ Mpc$^{-1}$. 

\begin{figure}[tb]
\begin{center}
\includegraphics[width=3.4in]{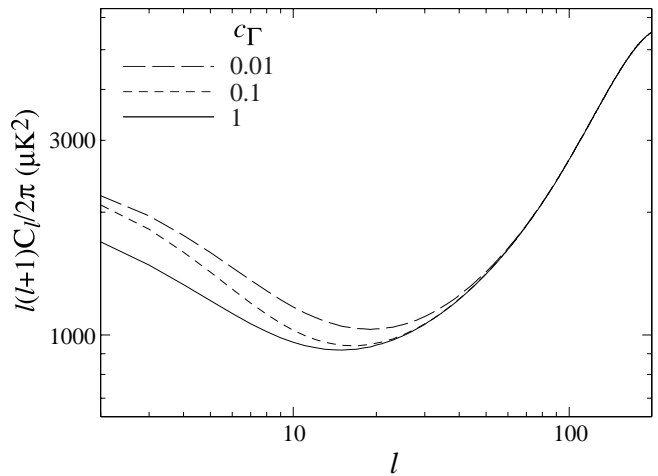}
\caption{Effect of changing the scale below which the quasistatic Poisson equation (\ref{eqn:qspoisson}) holds
through $c_\Gamma$.  At a fixed metric ratio $g$, the ISW effect is enhanced by lowering $c_{\Gamma}$
and delaying the onset of the quasistatic dynamics.  The other parameters have been
fixed to $g_{0}=1$, $c_{g}=0.01$ and $w_{e}=-1$.} \label{fig:gamma}
\end{center}
\end{figure}

Changing the quasistatic transition scale through $c_{\Gamma}$ has a substantial impact
on the relationship between the metric ratio $g$ and the ISW effect.  As discussed in \cite{HuSaw07b}, the superhorizon impact of $g$ implied by the conservation of the comoving curvature
$\zeta$ tends to be substantially larger than the
quasistatic limit would imply.
In Fig.~\ref{fig:gamma} we show the effect of decreasing $c_\Gamma$ for
$g_{0}=1$, $c_{g}=0.01$ and $w_{e}=-1$.  This extends
the region where the superhorizon behavior hold.   Once $c_{\Gamma}
\lesssim 0.1$, wavenumbers relevant to the ISW effect are impacted leading to 
a substantial enhancement of the effect for the same $g$.   The amount of enhancement
then depends also on the superhorizon parameter 
$f_\zeta$ and we take $f_\zeta=0.4 g_{\rm SH}$ throughout as
an illustrative example. For the ISW effect $f_\zeta$ is largely degenerate with $c_{\Gamma}$ since
they both control the interpolation between the superhorizon behavior and quasistatic
behavior both of which only depend on $g$ \cite{HuSaw07b}.

\begin{figure}[tb]
\begin{center}
\includegraphics[width=3.4in]{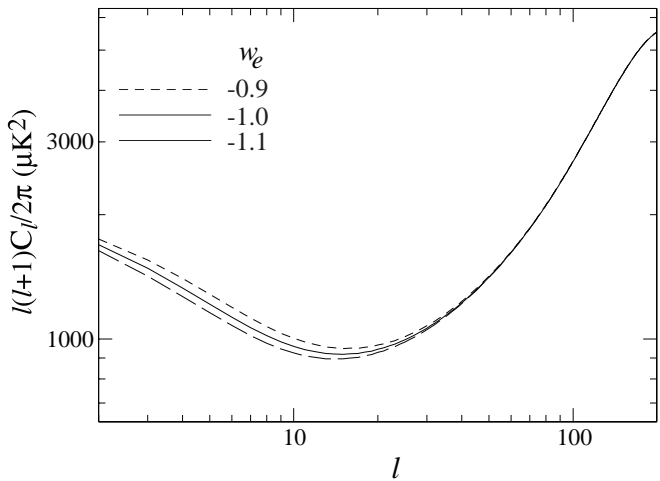}
\caption{Effect of changing the expansion history through $w_e$ is small compared with
cosmic variance and the effect of the PPF parameters for small deviations 
$-1.1 \le w_{e} \le -0.9$.   The other parameters have been fixed to
$g_{0}=1$, $c_{g}=0.01$ and $c_{\Gamma}=1$.} \label{fig:w}
\end{center}
\end{figure}
Finally, 
 in Fig.~\ref{fig:w} we show the effect of varying the background expansion history for
 a fixed $g_{0}=1$, $c_{g}=0.01$ and $c_{\Gamma}=1$.   Here $\Omega_{e}$ is adjusted
 to keep the distance to recombination fixed for the same $\Omega_{m}h^{2}$ and 
 $\Omega_{b}h^{2}$ of the fiducial model.
 For the relatively small variations allowed by  cosmological distance tests, 
 the changes are  small compared with cosmic variance and detectable variations in $g$.

\subsection{Constraints}
\label{sec:WMAP}

In the previous section, we have shown that for a given transition scale $c_{\Gamma}$,
the ISW effect constrains the metric ratio parameter $g$ at $k\sim 10^{-3}$ Mpc$^{{-1}}$
during the acceleration epoch.    We can now assess how well WMAP constrains
the amplitude of deviations in $g$.  In lieu of a joint analysis
of cosmological data sets that determine the distance redshift relation, we have here fixed the expansion
history and other cosmological parameters to $\Lambda$CDM.  Given how well most parameters
are currently fixed, the most important caveat introduced by this assumption is that
the initial power spectrum is taken to be of the power law form given by Eqn.~(\ref{eqn:pzeta}).  
We shall return to this point below.

\begin{figure}[tb]
\begin{center}
\includegraphics[width=3.4in]{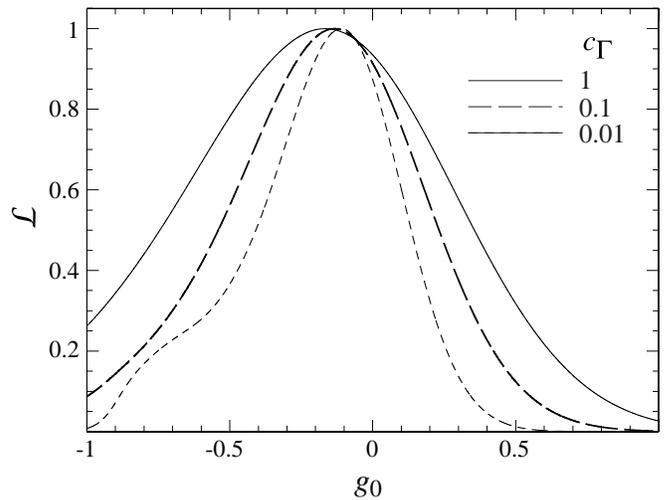}
\caption{WMAP likelihood of models as a function of $g_{0}$ for various
$c_{\Gamma}$ with $c_{g}=0.01$ and other parameters fixed.  Taking $c_{\Gamma}=1$ provides
conservative constraints. Normalization is arbitrary.} \label{fig:g0like}
\end{center}
\end{figure}

We begin by examining constraints on the  amplitude parameter $g_{0}
=g(\ln a=0)$ of 
Eqn.~(\ref{eqn:gevolution}). The relative WMAP likelihood is shown in Fig.~\ref{fig:g0like}
for several choices of $c_{\Gamma}$ and $c_{g}<0.01$.   Constraints tighten as $c_{\Gamma}$
decreases due to the enhancement in the ISW effect shown in Fig.~\ref{fig:gamma}.
A conservative interpretation of the constraint that is also well-motivated by the DGP and
$f(R)$ examples would be to take $c_{\Gamma} =1$.  The distributions are all consistent with
$g_{0}=0$. The statistically insignificant preference for negative $g_{0}$ is associated with a slight
lowering of the low $\ell$ multipoles favored by the data as we discuss in the
next section.  

Even taking the conservative $c_{\Gamma}=1$ limit, 
there is a problem in interpreting the constraint in a general context.  The parameter
 $g_{0}$ represents the metric ratio today and is not directly constrained
by the data.  We have furthermore assumed an ad hoc evolution of $g$ in Eqn.~(\ref{eqn:gevolution})
motivated only by a rough scaling with the impact on the expansion rate.

In order to place more general constraints on the metric ratio $g$,
we use a principal components construction to determine the temporal
weights that the ISW effect actually constrain.
The principal component construction begins with the calculation of the Fisher matrix
\begin{align}
F_{ij} = \sum_\ell (\ell + 1/2) {\partial \ln C_\ell \over \partial p_i}{\partial \ln C_\ell \over \partial p_j}\,,
\end{align}
where $p_i$ are a set of $N_{p}$ parameters that represent the values of a spline interpolated
function $p(\ln a)$
\begin{align}
g_{\rm SH}(\ln a) = p(\ln a) \left( {\rho_e  \over \rho_T} {\Omega_T \over \Omega_e} \right)^{1/2}\,,
\end{align}
evaluated at equally spaced intervals in $\ln a$. 
The derivatives are evaluated at the $\Lambda$CDM model of $p_{i}=0$.
We further take $c_{g}<0.01$ and $c_{\Gamma}=1$ here.

We then decompose the Fisher matrix into principal components indexed by $\mu$
\begin{equation}
F_{ij} = \sum_{\mu} S_{i\mu} \sigma_{\mu}^{-2} S_{j \mu}\,,
\end{equation}
where $\sigma_{\mu}^{2}$ is the Fisher estimate of the variance of mode associated with the
linear combination $\sum_{i} S_{i\mu} p_{i}$.  Rank ordered in increasing variance, we find
that the second mode has 5 times the variance of the first mode and so we shall use only the
first mode in the following analysis.

\begin{figure}[tb]
\begin{center}
\includegraphics[width=3.4in]{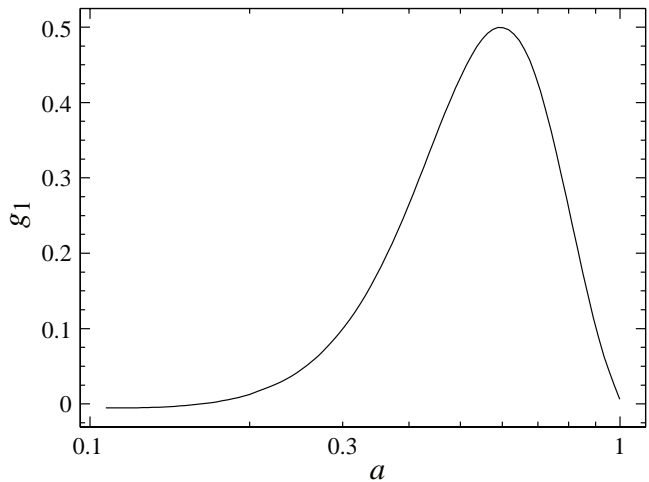}
\caption{First principal component $g_{1}$ of the metric ratio parameter as a function of
scale factor $a$.  Constraints derived from this component represent a temporal average of
$g$  of this functional form through Eqn.~(\ref{eqn:gprojection}).  Note that the maximum value $g_{\rm max}=0.5$ at $z=0.68$.} \label{fig:plotfirst}
\end{center}
\end{figure}

More specifically, let us define a continuous function $V_{1}$ which is the high $N_{p}$ limit of
the discrete eigenmode
\begin{equation}
V_1(\ln a_i) \propto S_{i1} \,,
\end{equation}
normalized so that  
\begin{equation}
\int d\ln a V_1^2 = 1 \,.
\end{equation}
The metric ratio $g$ that this mode represents is given by
\begin{equation}
g_1(\ln a) = \left( {\rho_e  \over \rho_T} {\Omega_T \over \Omega_e} \right)^{1/2} V_1 (\ln a)
\end{equation}
and is shown in Fig.~\ref{fig:plotfirst}.  Note that the maximum value attained in this
function is $g_{\rm max} = 0.5$ at a redshift of $z_{\rm max}=0.68$.  The ISW effect thus
constrains a weighted average of $g$ around $z_{\rm max}$.  The falloff in sensitivity 
near the present epoch is due to projection effects.

\begin{figure}[tb]
\begin{center}
\includegraphics[width=3.4in]{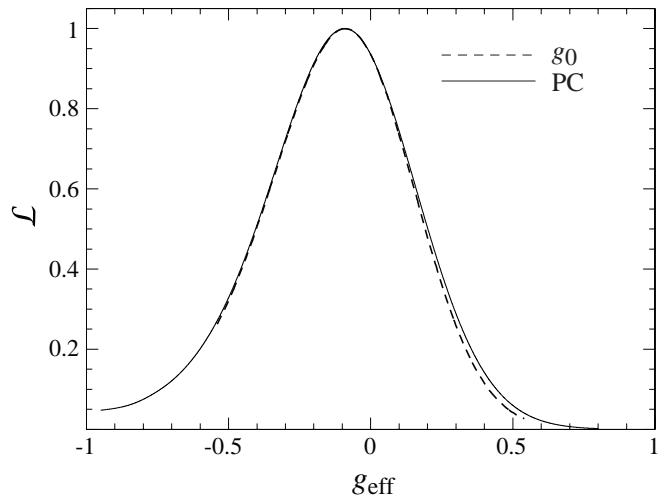}
\caption{WMAP constraints on $g_{\rm eff}$ from the first principal component.  Note that
$g_{\rm eff}$ is the maximum value of $g$ attained by the principal component.  For comparison
the direct constraint on $g_{0}$ with $c_{\Gamma}=0.1$ from Fig.~\ref{fig:g0like} is overplotted
with the
conversion $g_{\rm eff}=0.54g_{0}$ from 
Eqn.~(\ref{eqn:gprojection}).} \label{fig:plotlike}
\end{center}
\end{figure}

We now use this functional form of $g$ to evaluate the WMAP constraint on the amplitude
$g_{\rm eff}$
\begin{equation}
g_{\rm SH}(\ln a)  ={ g_{\rm eff} {g_1(\ln a) \over g_{\rm max} }} \,.
\end{equation}
   Fig.~\ref{fig:plotlike}
shows that the posterior probability distribution of $g_{\rm eff}$ is nearly Gaussian
and that 
\begin{equation}
g_{\rm eff}=-0.12\pm 0.27 \,.  
\end{equation}
Note that $g_{\rm eff}$ is the value of $g_{\rm SH}$ at the $z_{\rm max}$.

Given a specific model for $g$, this constraint should be interpreted as one on 
a given weighted average of $g$ around $z_{\rm max}$;  specifically\begin{equation}
g_{\rm eff} =   
g_{\rm max}
\int d\ln a p(\ln a) V_1(\ln a) \,.
\label{eqn:gprojection}
\end{equation}
For example, employing the model of Eqn.~(\ref{eqn:gevolution}) we obtain
$g_{\rm eff} = 0.54 g_{0}$.  In Fig.~\ref{fig:plotlike} we show that the direct constraints
on $g_{0}$ compare well with those inferred from $g_{\rm eff}$.

\subsection{Designer Models and Caveats}
\label{sec:designer}

The constraints obtained in the previous section are fairly general in that
a model that violates them while maintaining the underlying assumptions of a smooth
initial power spectrum and an expansion history close to $\Lambda$CDM will be
disfavored by the data.   However there remains a wide range of interesting phenomenology associated
with acceptable models and a weakening of the underlying assumptions.

\begin{figure}
\begin{center}
\includegraphics[width=3.4in]{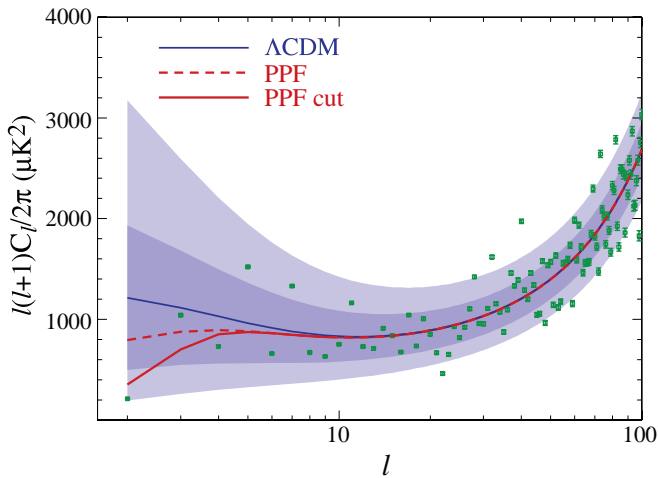}
\caption{Temperature power spectrum of the $\Lambda$CDM model with the WMAP data
and noise error bars overplotted.   Shaded regions represent the 68\% and 95\% cosmic 
variance intervals for the model.  A PPF model constructed to eliminate the ISW effect
can lower the power in the quadrupole.  Shown is the model discussed in the text with and
without an initial power spectrum cut off at $k_{\rm min}=2.4 \times 10^{-4}$Mpc$^{-1}$.} \label{fig:cl}
\end{center}
\end{figure}

As an example of phenomenology that PPF models can produce that is not possible in $\Lambda$CDM,
consider the statistical curiosity of the low quadrupole power in the CMB.  
Figure~\ref{fig:cl} shows that in the maximum likelihood $\Lambda$CDM model, the observed
quadrupole is near the $95\%$ cosmic variance limit.   A quadrupole realization as
extreme as the WMAP data only occurs in $\sim 5.5$\% of the time.  Of course in the first say 100 multipoles 
one expects and finds a few such events. The quadrupole is only special because it represents the largest
observable scale and may hint at new physics beyond the standard cosmological model.

Construction of a model that sharply lowers the power in the quadrupole but not the octopole
is hindered by the fact that in $\Lambda$CDM it gets nearly equal contributions from
the ISW effect and the Sachs-Wolfe effect.  These contributions are separated by a 
decade in physical scale making it difficult to invoke initial conditions
as an explanation of the low quadrupole \cite{ConPelKofLin03}.  Furthermore for dynamical dark energy models
based on scalar fields, it is difficult to reduce the ISW effect while maintaining an
expansion history close to $\Lambda$CDM  \cite{BeaDor04,GorHu04}.

As we have seen, a modified gravity model can alter the evolution of the metric
even for an expansion history that is indistinguishable from $\Lambda$CDM.  Let us
then construct a model which nearly eliminates the ISW effect at low multipoles
while leaving the spectrum at $\ell \gtrsim 10$ unchanged.  To enhance the impact on the
low multipoles let us take $c_\Gamma=0.1$, $c_g < 0.01$
and take a very steep evolution of $g$ such
that deviations appear only recently
\begin{align}
g_{\rm SH}(\ln a) = A \left( {\rho_e  \over \rho_T} {\Omega_T \over \Omega_e} \right)^{2} \,.
\label{eqn:gdesignerevolution}
\end{align}
We find that $A=-0.7$ causes a growth at late times that nearly offsets the
earlier decay leaving the ISW effect nearly absent in the quadrupole.   Because of the steep
evolution, this model evades the constraints from the previous section. 
 We show the spectrum
of this model in Fig.~\ref{fig:cl} (dashed line).  This model is a better fit to the WMAP data
with a $\Delta 2\ln L = -1.5$.  Of course given the number of parameters and ad hoc assumptions
of this model, such an improvement is not statistically significant.  On the other hand, explorations
of this sort can suggest what types of modified gravity models would be fruitful to study: in
this case, models
with $g<0$ on the horizon scale near the present. 

With the ISW effect nearly eliminated the quadrupole can be further lowered by a change in the initial
power spectrum of Eqn.~(\ref{eqn:pzeta}).  Placing a cutoff of $k_{\rm min}=2.4\times 10^{-4}$ Mpc$^{-1}$
produces the spectrum shown in Fig.~\ref{fig:cl} and an improved likelihood of $\Delta 2 \ln L = -2.8$.

This example also highlights the primary caveat of applying the constraint on $g_{\rm eff}$ in the previous
section.  That constraint assumes a featureless initial power spectrum where the constraints at high $\ell$
nearly fix the initial power at the horizon scale today.  While this is reasonable given the expectation that
gravity is only modified at late times, it remains an assumption nonetheless.

\subsection{Discussion}
\label{sec:discussion}
 
 We have generalized the parametrized post-Friedmann (PPF) description of cosmic acceleration from
 modified gravity to include multiple relativistic matter species and spatial curvature.  This generalization 
 facilitates the adaptation of Einstein-Boltzmann codes for modified gravity.  We have adapted a comoving
 gauge code to study constraints on  deviations from general relativity on the largest scales through
 the CMB. 
 
 The integrated Sachs-Wolfe (ISW) effect probes the evolution of the metric $\Phi_{-}$ on scales
 near the horizon during the acceleration
 epoch.    Under the PPF parametrization, modifications to gravity can change the amplitude
 and shape of the ISW contributions to the low multipole CMB temperature power spectrum
 even if the expansion history is indistinguishable from that of a cosmological constant.   For example
 in some regions of parameter space, the ISW effect can be nearly eliminated at the quadrupole
 bringing the predicted ensemble average quadrupole nearer to its observed value on our sky. 
 
 We use the WMAP data to constrain such modifications and find that deviations in the metric
 ratio parameter $g$ are constrained at the level of $g_{\rm eff}=-0.12 \pm 0.27$  at $z \sim 0.7$
 and $k \sim 10^{-3}$ Mpc$^{-1}$.
 Correspondingly, specific models such as the DGP  and some $f(R)$ models are 
 constrained by the data at levels consistent with the findings of previous work 
 (e.g. \cite{SonSawHu06,SonPeiHu07}).  By phrasing these and other constraints in the model-independent
 PPF description, we gain a more general understanding of what aspects of general relativity are tested
 by the observations.
 
 The modified gravity framework studied here should also enable future studies of other
 cosmological observables on large scales that are beyond the scope of this work such as the lensing of the CMB and faint galaxies
 as well as CMB-galaxy correlations.

\appendix

\section{Synchronous Gauge}

Most public Boltzmann codes are written in the synchronous gauge (e.g. \cite{SelZal96, Lewetal00}).  For completeness,
we outline here an implementation of PPF that is appropriate for such codes.

The synchronous gauge is defined by the conditions
\begin{align}
	 A &=  B = 0\,, \nonumber \\
	{\eta_T} &\equiv -\frac{1}{3}  H_T - H_{L}\,, \nonumber\\
  h_{L} &= 6H_L\,, \nonumber\\
	T &= a^{-1} \int d\eta a \tilde A + c_1 a^{-1} \nonumber\,, \\
	L &= - \int d\eta(\tilde B+kT) + c_2 \,.
\end{align}
which leave two constants $c_1$ and $c_2$ to be specified by the initial choice
of time slicing and threading (e.g. the rest frame of the dark matter).  
On the other  hand the gauge transformation from a specific synchronous gauge
choice to the comoving gauge used in the main paper is completely defined
\begin{eqnarray}
T &=& (v_T^ s/k)\,, \nonumber\\
L &=& {3 \over k} (\eta_T + {1 \over 6} h_L ) \,,
\end{eqnarray}
where the superscript $s$ denotes synchronous gauge quantities.
Note further that the comoving gauge variables depend only on $L'$ and not $L$
since $H_T$ vanishes in the comoving gauge.

The gauge transformation equations in the main text allow for the evolution of the dark energy 
parameter $\Gamma$ to be expressed in terms of the synchronous gauge variables.  
We begin by expressing the comoving gauge density fluctuations as 
\begin{eqnarray}
\Delta_i \rho_i &=& \delta_i^s \rho_i - \rho_i' (v_T^s/k_{H}) \,. 
\end{eqnarray}

This allows us to obtain $\Phi_-$ in terms of $\delta_i^s$, $v_i^s$ and $\Gamma$.
The sources for $\Gamma$ must also be expressed in terms of synchronous gauge quantities.
In addition to $\Phi_-$ and the gauge invariant $\pi_i$ there is $V_T$.
  Starting with the 
gauge relation (\ref{eqn:comnewt1})
\begin{equation}
V_{T} = k_{H}(\Phi-\zeta)\,,
\end{equation}
we can then use the synchronous to comoving gauge relation
\begin{equation}
\zeta = -\eta_{T} - {v_{T}^{s}\over k_{H}}\,,
\end{equation}
and the metric relation implied by Eqn.~(\ref{eqn:gdef})
\begin{equation}
\Phi = (g+1)\Phi_{-} - {4 \pi G \over H^{2}k_{H}^{2}}p_T\Pi_T
\end{equation}
to obtain
\begin{equation}
{V_{T} \over k_{H}} = {v_{T}^{s} \over k_{H}} + (g+1)\Phi_{-} - {4 \pi G \over H^{2}k_{H}^{2}} p_T\Pi_T +\eta_{T} \,,
\end{equation}
which now gives the evolution equation (\ref{eqn:gammaeom}) 
for $\Gamma$ in terms of synchronous variables only.

The conservation equations for the separate components of real matter and radiation are of  
course unaltered but require the evolution of $h_T'$ and $\eta_T$ from the Einstein equations.
These are modified to include the stress energy of the effective dark energy.
 We can use the
fluid correspondence and gauge relations to obtain
\begin{equation}
\delta_{e}^{s} = \Delta_{e} - 3(1+w_{e}){v_{T}^{s} \over k_{H}} \,.
\end{equation}
Note that 
\begin{equation}
v_{e}^{s} = V_{e}-V_T + v_T^s \,,
\end{equation}
and is given by Eqn.~(\ref{eqn:veeff}) in terms of $\Phi_-$, $\Gamma$, and $V_T$. 
Since $p_{e}\Pi_{e}$ is gauge invariant, the remaining component is the pressure fluctuation
obtained from
\begin{equation}
\delta p_e^s = \Delta p_e + p_e' {v_T^s \over k_H} \,.
\end{equation}

The third and fourth Einstein equations (\ref{eqn:einstein}) give 
\begin{align}
&\eta_{T}' - {3K \over k^{2}} (\eta_{T}' + {1 \over 6} h_L') \nonumber\\
&\qquad = {4\pi G \over H^2} \left[
(\rho_{T}+p_{T}){v_{T}^s\over k_H} + (\rho_{e}+p_{e}){v_{e}^s\over k_H} \right] \,,
\end{align}
and
\begin{align}
& h_{L}'' + \left( 2 + {H'\over H} \right) h_{L}' \nonumber\\
&\qquad= -{8\pi G \over H^2}\left[ \delta_{T}^{s}\rho_{T}+ 3\delta p_{T}^{s}
+ \delta_{e}^{s}\rho_{T}+ 3\delta p_{e}^{s} \right] \,.
\label{eqn:hLeom}
\end{align}
Since the pressure perturbation $\delta p_e^{s}$ is cumbersome to evaluate, 
one can alternately replace equation (\ref{eqn:hLeom}) with the first Einstein
equation (\ref{eqn:einstein}) 
\begin{align}
h_L' = 2 \ck k_H^2 \eta_T + {8\pi G \over H^2} (\delta_{T}^{s}\rho_{T}^{s} + \delta_{e}^{s}\rho_{e}^{s} 
)\,,
\end{align}
which involves $v_e^s$ and $\delta_e \rho_e^s$ only.  The latter system is the set of
Einstein equations solved by CAMB \cite{Lewetal00}.  

The only drawback of using the latter
system is
that the matter density and metric $h_L$ depend to leading order on the velocities
of the effective (or real) dark energy through $\eta_T$ even at $k_H\gg 1$ since the synchronous initial
conditions typically set
the dark matter velocity to zero.  For example in ordinary smooth dark energy models,
dark energy perturbations cannot be consistently set to zero in this set of equations when
solving for the matter power spectrum
whereas they may be in the former.

\vspace{1.cm}

\noindent {\it Acknowledgments}: I thank Ignacy Sawicki and
Yong-Seon Song for useful conversations and Wenjuan Fang for pointing out several typos in an earlier version.
This work was supported by the
U.S.~Dept.~of Energy contract DE-FG02-90ER-40560, 
the David and Lucile Packard Foundation
and the KICP under NSF PHY-0114422. 
\hfill\vfill
\bibliography{Hu08a}

\begin{thebibliography}{35}
\expandafter\ifx\csname natexlab\endcsname\relax\def\natexlab#1{#1}\fi
\expandafter\ifx\csname bibnamefont\endcsname\relax
  \def\bibnamefont#1{#1}\fi
\expandafter\ifx\csname bibfnamefont\endcsname\relax
  \def\bibfnamefont#1{#1}\fi
\expandafter\ifx\csname citenamefont\endcsname\relax
  \def\citenamefont#1{#1}\fi
\expandafter\ifx\csname url\endcsname\relax
  \def\url#1{\texttt{#1}}\fi
\expandafter\ifx\csname urlprefix\endcsname\relax\def\urlprefix{URL }\fi
\providecommand{\bibinfo}[2]{#2}
\providecommand{\eprint}[2][]{\url{#2}}

\bibitem[{\citenamefont{Ishak et~al.}(2006)\citenamefont{Ishak, Upadhye, and
  Spergel}}]{IshUpaSpe06}
\bibinfo{author}{\bibfnamefont{M.}~\bibnamefont{Ishak}},
  \bibinfo{author}{\bibfnamefont{A.}~\bibnamefont{Upadhye}}, \bibnamefont{and}
  \bibinfo{author}{\bibfnamefont{D.~N.} \bibnamefont{Spergel}},
  \bibinfo{journal}{Phys. Rev.} \textbf{\bibinfo{volume}{D74}},
  \bibinfo{pages}{043513} (\bibinfo{year}{2006}), \eprint{astro-ph/0507184}.

\bibitem[{\citenamefont{Knox et~al.}(2006)\citenamefont{Knox, Song, and
  Tyson}}]{KnoSonTys06}
\bibinfo{author}{\bibfnamefont{L.}~\bibnamefont{Knox}},
  \bibinfo{author}{\bibfnamefont{Y.-S.} \bibnamefont{Song}}, \bibnamefont{and}
  \bibinfo{author}{\bibfnamefont{J.~A.} \bibnamefont{Tyson}},
  \bibinfo{journal}{Phys. Rev.} \textbf{\bibinfo{volume}{D74}},
  \bibinfo{pages}{023512} (\bibinfo{year}{2006}).

\bibitem[{\citenamefont{Wang et~al.}(2007)\citenamefont{Wang, Hui, May, and
  Haiman}}]{WanHuiMayHai07}
\bibinfo{author}{\bibfnamefont{S.}~\bibnamefont{Wang}},
  \bibinfo{author}{\bibfnamefont{L.}~\bibnamefont{Hui}},
  \bibinfo{author}{\bibfnamefont{M.}~\bibnamefont{May}}, \bibnamefont{and}
  \bibinfo{author}{\bibfnamefont{Z.}~\bibnamefont{Haiman}}
  (\bibinfo{year}{2007}), \eprint{arXiv:0705.0165 [astro-ph]}.

\bibitem[{\citenamefont{Song}(2006)}]{Son06}
\bibinfo{author}{\bibfnamefont{Y.-S.} \bibnamefont{Song}}
  (\bibinfo{year}{2006}), \eprint{astro-ph/0602598}.

\bibitem[{\citenamefont{Huterer and Linder}(2007)}]{HutLin07}
\bibinfo{author}{\bibfnamefont{D.}~\bibnamefont{Huterer}} \bibnamefont{and}
  \bibinfo{author}{\bibfnamefont{E.~V.} \bibnamefont{Linder}},
  \bibinfo{journal}{Phys. Rev.} \textbf{\bibinfo{volume}{D75}},
  \bibinfo{pages}{023519} (\bibinfo{year}{2007}), \eprint{astro-ph/0608681}.

\bibitem[{\citenamefont{Caldwell et~al.}(2007)\citenamefont{Caldwell, Cooray,
  and Melchiorri}}]{CalCooMel07}
\bibinfo{author}{\bibfnamefont{R.}~\bibnamefont{Caldwell}},
  \bibinfo{author}{\bibfnamefont{A.}~\bibnamefont{Cooray}}, \bibnamefont{and}
  \bibinfo{author}{\bibfnamefont{A.}~\bibnamefont{Melchiorri}},
  \bibinfo{journal}{Phys. Rev.} \textbf{\bibinfo{volume}{D76}},
  \bibinfo{pages}{023507} (\bibinfo{year}{2007}), \eprint{astro-ph/0703375}.

\bibitem[{\citenamefont{Amendola et~al.}(2007)\citenamefont{Amendola, Kunz, and
  Sapone}}]{Amendola:2007rr}
\bibinfo{author}{\bibfnamefont{L.}~\bibnamefont{Amendola}},
  \bibinfo{author}{\bibfnamefont{M.}~\bibnamefont{Kunz}}, \bibnamefont{and}
  \bibinfo{author}{\bibfnamefont{D.}~\bibnamefont{Sapone}}
  (\bibinfo{year}{2007}), \eprint{arXiv:0704.2421 [astro-ph]}.

\bibitem[{\citenamefont{{Zhang} et~al.}(2007)\citenamefont{{Zhang}, {Liguori},
  {Bean}, and {Dodelson}}}]{ZhaLigBeaDod07}
\bibinfo{author}{\bibfnamefont{P.}~\bibnamefont{{Zhang}}},
  \bibinfo{author}{\bibfnamefont{M.}~\bibnamefont{{Liguori}}},
  \bibinfo{author}{\bibfnamefont{R.}~\bibnamefont{{Bean}}}, \bibnamefont{and}
  \bibinfo{author}{\bibfnamefont{S.}~\bibnamefont{{Dodelson}}},
  \bibinfo{journal}{ArXiv e-prints} \textbf{\bibinfo{volume}{704}}
  (\bibinfo{year}{2007}), \eprint{0704.1932}.

\bibitem[{\citenamefont{Amin et~al.}(2007)\citenamefont{Amin, Wagoner, and
  Blandford}}]{AmiWagBla07}
\bibinfo{author}{\bibfnamefont{M.~A.} \bibnamefont{Amin}},
  \bibinfo{author}{\bibfnamefont{R.~V.} \bibnamefont{Wagoner}},
  \bibnamefont{and} \bibinfo{author}{\bibfnamefont{R.~D.}
  \bibnamefont{Blandford}} (\bibinfo{year}{2007}), \eprint{arXiv:0708.1793
  [astro-ph]}.

\bibitem[{\citenamefont{Jain and Zhang}(2007)}]{Jain:2007yk}
\bibinfo{author}{\bibfnamefont{B.}~\bibnamefont{Jain}} \bibnamefont{and}
  \bibinfo{author}{\bibfnamefont{P.}~\bibnamefont{Zhang}}
  (\bibinfo{year}{2007}), \eprint{arXiv:0709.2375 [astro-ph]}.

\bibitem[{\citenamefont{Hu and Sawicki}(2007)}]{HuSaw07b}
\bibinfo{author}{\bibfnamefont{W.}~\bibnamefont{Hu}} \bibnamefont{and}
  \bibinfo{author}{\bibfnamefont{I.}~\bibnamefont{Sawicki}},
  \bibinfo{journal}{Phys. Rev.} \textbf{\bibinfo{volume}{D76}},
  \bibinfo{pages}{104043} (\bibinfo{year}{2007}), \eprint{arXiv:0708.1190
  [astro-ph]}.

\bibitem[{\citenamefont{Dvali et~al.}(2000)\citenamefont{Dvali, Gabadadze, and
  Porrati}}]{DvaGabPor00}
\bibinfo{author}{\bibfnamefont{G.~R.} \bibnamefont{Dvali}},
  \bibinfo{author}{\bibfnamefont{G.}~\bibnamefont{Gabadadze}},
  \bibnamefont{and} \bibinfo{author}{\bibfnamefont{M.}~\bibnamefont{Porrati}},
  \bibinfo{journal}{Phys. Lett.} \textbf{\bibinfo{volume}{B485}},
  \bibinfo{pages}{208} (\bibinfo{year}{2000}), \eprint{hep-th/0005016}.

\bibitem[{\citenamefont{Carroll et~al.}(2004)\citenamefont{Carroll, Duvvuri,
  Trodden, and Turner}}]{Caretal03}
\bibinfo{author}{\bibfnamefont{S.~M.} \bibnamefont{Carroll}},
  \bibinfo{author}{\bibfnamefont{V.}~\bibnamefont{Duvvuri}},
  \bibinfo{author}{\bibfnamefont{M.}~\bibnamefont{Trodden}}, \bibnamefont{and}
  \bibinfo{author}{\bibfnamefont{M.~S.} \bibnamefont{Turner}},
  \bibinfo{journal}{Phys. Rev.} \textbf{\bibinfo{volume}{D70}},
  \bibinfo{pages}{043528} (\bibinfo{year}{2004}), \eprint{astro-ph/0306438}.

\bibitem[{\citenamefont{Nojiri and Odintsov}(2003)}]{NojOdi03}
\bibinfo{author}{\bibfnamefont{S.}~\bibnamefont{Nojiri}} \bibnamefont{and}
  \bibinfo{author}{\bibfnamefont{S.~D.} \bibnamefont{Odintsov}},
  \bibinfo{journal}{Phys. Rev.} \textbf{\bibinfo{volume}{D68}},
  \bibinfo{pages}{123512} (\bibinfo{year}{2003}), \eprint{hep-th/0307288}.

\bibitem[{\citenamefont{Capozziello et~al.}(2003)\citenamefont{Capozziello,
  Carloni, and Troisi}}]{Capozziello:2003tk}
\bibinfo{author}{\bibfnamefont{S.}~\bibnamefont{Capozziello}},
  \bibinfo{author}{\bibfnamefont{S.}~\bibnamefont{Carloni}}, \bibnamefont{and}
  \bibinfo{author}{\bibfnamefont{A.}~\bibnamefont{Troisi}}
  (\bibinfo{year}{2003}), \eprint{astro-ph/0303041}.

\bibitem[{\citenamefont{{Seljak} and {Zaldarriaga}}(1996)}]{SelZal96}
\bibinfo{author}{\bibfnamefont{U.}~\bibnamefont{{Seljak}}} \bibnamefont{and}
  \bibinfo{author}{\bibfnamefont{M.}~\bibnamefont{{Zaldarriaga}}},
  \bibinfo{journal}{\apj} \textbf{\bibinfo{volume}{469}}, \bibinfo{pages}{437}
  (\bibinfo{year}{1996}), \eprint{astro-ph/9603033}.

\bibitem[{\citenamefont{{Lewis} et~al.}(2000)\citenamefont{{Lewis},
  {Challinor}, and {Lasenby}}}]{Lewetal00}
\bibinfo{author}{\bibfnamefont{A.}~\bibnamefont{{Lewis}}},
  \bibinfo{author}{\bibfnamefont{A.}~\bibnamefont{{Challinor}}},
  \bibnamefont{and}
  \bibinfo{author}{\bibfnamefont{A.}~\bibnamefont{{Lasenby}}},
  \bibinfo{journal}{\apj} \textbf{\bibinfo{volume}{538}}, \bibinfo{pages}{473}
  (\bibinfo{year}{2000}), \eprint{astro-ph/9911177}.

\bibitem[{\citenamefont{{Song} et~al.}(2006)\citenamefont{{Song}, {Sawicki},
  and {Hu}}}]{SonSawHu06}
\bibinfo{author}{\bibfnamefont{Y.}~\bibnamefont{{Song}}},
  \bibinfo{author}{\bibfnamefont{I.}~\bibnamefont{{Sawicki}}},
  \bibnamefont{and} \bibinfo{author}{\bibfnamefont{W.}~\bibnamefont{{Hu}}},
  \bibinfo{journal}{\prd} \textbf{\bibinfo{volume}{75}},
  \bibinfo{pages}{064003} (\bibinfo{year}{2006}), \eprint{astro-ph/0606286}.

\bibitem[{\citenamefont{Lue et~al.}(2004)\citenamefont{Lue, Scoccimarro, and
  Starkman}}]{LueScoSta04}
\bibinfo{author}{\bibfnamefont{A.}~\bibnamefont{Lue}},
  \bibinfo{author}{\bibfnamefont{R.}~\bibnamefont{Scoccimarro}},
  \bibnamefont{and} \bibinfo{author}{\bibfnamefont{G.~D.}
  \bibnamefont{Starkman}}, \bibinfo{journal}{Phys. Rev.}
  \textbf{\bibinfo{volume}{D69}}, \bibinfo{pages}{124015}
  (\bibinfo{year}{2004}), \eprint{astro-ph/0401515}.

\bibitem[{\citenamefont{{Song} et~al.}(2007)\citenamefont{{Song}, {Peiris}, and
  {Hu}}}]{SonPeiHu07}
\bibinfo{author}{\bibfnamefont{Y.}~\bibnamefont{{Song}}},
  \bibinfo{author}{\bibfnamefont{H.}~\bibnamefont{{Peiris}}}, \bibnamefont{and}
  \bibinfo{author}{\bibfnamefont{W.}~\bibnamefont{{Hu}}},
  \bibinfo{journal}{\prd} \textbf{\bibinfo{volume}{76}},
  \bibinfo{pages}{063517} (\bibinfo{year}{2007}), \eprint{0706.2399
  [astro-ph]}.

\bibitem[{\citenamefont{Schmidt et~al.}(2007)\citenamefont{Schmidt, Liguori,
  and Dodelson}}]{Schmidt:2007vj}
\bibinfo{author}{\bibfnamefont{F.}~\bibnamefont{Schmidt}},
  \bibinfo{author}{\bibfnamefont{M.}~\bibnamefont{Liguori}}, \bibnamefont{and}
  \bibinfo{author}{\bibfnamefont{S.}~\bibnamefont{Dodelson}},
  \bibinfo{journal}{Phys. Rev.} \textbf{\bibinfo{volume}{D76}},
  \bibinfo{pages}{083518} (\bibinfo{year}{2007}), \eprint{arXiv:0706.1775
  [astro-ph]}.

\bibitem[{\citenamefont{Pogosian and Silvestri}(2008)}]{Pogosian:2007sw}
\bibinfo{author}{\bibfnamefont{L.}~\bibnamefont{Pogosian}} \bibnamefont{and}
  \bibinfo{author}{\bibfnamefont{A.}~\bibnamefont{Silvestri}},
  \bibinfo{journal}{Phys. Rev.} \textbf{\bibinfo{volume}{D77}},
  \bibinfo{pages}{023503} (\bibinfo{year}{2008}), \eprint{0709.0296}.

\bibitem[{\citenamefont{Kunz and Sapone}(2007)}]{Kunz:2006ca}
\bibinfo{author}{\bibfnamefont{M.}~\bibnamefont{Kunz}} \bibnamefont{and}
  \bibinfo{author}{\bibfnamefont{D.}~\bibnamefont{Sapone}},
  \bibinfo{journal}{Phys. Rev. Lett.} \textbf{\bibinfo{volume}{98}},
  \bibinfo{pages}{121301} (\bibinfo{year}{2007}), \eprint{astro-ph/0612452}.

\bibitem[{\citenamefont{Bashinsky}(2007)}]{Bashinsky:2007yc}
\bibinfo{author}{\bibfnamefont{S.}~\bibnamefont{Bashinsky}}
  (\bibinfo{year}{2007}), \eprint{arXiv:0707.0692 [astro-ph]}.

\bibitem[{\citenamefont{{Hu}}(1998)}]{Hu98}
\bibinfo{author}{\bibfnamefont{W.}~\bibnamefont{{Hu}}}, \bibinfo{journal}{\apj}
  \textbf{\bibinfo{volume}{506}}, \bibinfo{pages}{485} (\bibinfo{year}{1998}),
  \eprint{astro-ph/9801234}.

\bibitem[{\citenamefont{{Bardeen}}(1980)}]{Bar80}
\bibinfo{author}{\bibfnamefont{J.~M.} \bibnamefont{{Bardeen}}},
  \bibinfo{journal}{\prd} \textbf{\bibinfo{volume}{22}}, \bibinfo{pages}{1882}
  (\bibinfo{year}{1980}).

\bibitem[{\citenamefont{{Kodama} and {Sasaki}}(1984)}]{KodSas84}
\bibinfo{author}{\bibfnamefont{H.}~\bibnamefont{{Kodama}}} \bibnamefont{and}
  \bibinfo{author}{\bibfnamefont{M.}~\bibnamefont{{Sasaki}}},
  \bibinfo{journal}{Prog. Theor. Phys. Suppl.} \textbf{\bibinfo{volume}{78}},
  \bibinfo{pages}{1} (\bibinfo{year}{1984}).

\bibitem[{\citenamefont{{Hu} and {Eisenstein}}(1999)}]{HuEis99}
\bibinfo{author}{\bibfnamefont{W.}~\bibnamefont{{Hu}}} \bibnamefont{and}
  \bibinfo{author}{\bibfnamefont{D.~J.} \bibnamefont{{Eisenstein}}},
  \bibinfo{journal}{\prd} \textbf{\bibinfo{volume}{59}},
  \bibinfo{pages}{083509} (\bibinfo{year}{1999}), \eprint{astro-ph/9809368}.

\bibitem[{\citenamefont{Bertschinger}(2006)}]{Ber06}
\bibinfo{author}{\bibfnamefont{E.}~\bibnamefont{Bertschinger}},
  \bibinfo{journal}{Astrophys. J.} \textbf{\bibinfo{volume}{648}},
  \bibinfo{pages}{797} (\bibinfo{year}{2006}), \eprint{astro-ph/0604485}.

\bibitem[{\citenamefont{{Hu} and {Okamoto}}(2003)}]{HuOka03}
\bibinfo{author}{\bibfnamefont{W.}~\bibnamefont{{Hu}}} \bibnamefont{and}
  \bibinfo{author}{\bibfnamefont{T.}~\bibnamefont{{Okamoto}}},
  \bibinfo{journal}{\prd} \textbf{\bibinfo{volume}{69}},
  \bibinfo{pages}{043004} (\bibinfo{year}{2003}), \eprint{astro-ph/0308049}.

\bibitem[{\citenamefont{{Hu} et~al.}(1998)\citenamefont{{Hu}, {Seljak},
  {White}, and {Zaldarriaga}}}]{HuSelWhiZal98}
\bibinfo{author}{\bibfnamefont{W.}~\bibnamefont{{Hu}}},
  \bibinfo{author}{\bibfnamefont{U.}~\bibnamefont{{Seljak}}},
  \bibinfo{author}{\bibfnamefont{M.}~\bibnamefont{{White}}}, \bibnamefont{and}
  \bibinfo{author}{\bibfnamefont{M.}~\bibnamefont{{Zaldarriaga}}},
  \bibinfo{journal}{\prd} \textbf{\bibinfo{volume}{57}}, \bibinfo{pages}{3290}
  (\bibinfo{year}{1998}), \eprint{astro-ph/9709066}.

\bibitem[{\citenamefont{Spergel et~al.}(2007)}]{Speetal06}
\bibinfo{author}{\bibfnamefont{D.~N.} \bibnamefont{Spergel}}
  \bibnamefont{et~al.} (\bibinfo{collaboration}{WMAP}),
  \bibinfo{journal}{Astrophys. J. Suppl.} \textbf{\bibinfo{volume}{170}},
  \bibinfo{pages}{377} (\bibinfo{year}{2007}), \eprint{astro-ph/0603449}.

\bibitem[{\citenamefont{{Contaldi} et~al.}(2003)\citenamefont{{Contaldi},
  {Peloso}, {Kofman}, and {Linde}}}]{ConPelKofLin03}
\bibinfo{author}{\bibfnamefont{C.}~\bibnamefont{{Contaldi}}},
  \bibinfo{author}{\bibfnamefont{M.}~\bibnamefont{{Peloso}}},
  \bibinfo{author}{\bibfnamefont{L.}~\bibnamefont{{Kofman}}}, \bibnamefont{and}
  \bibinfo{author}{\bibfnamefont{A.}~\bibnamefont{{Linde}}},
  \bibinfo{journal}{JCAP} \textbf{\bibinfo{volume}{0307}}, \bibinfo{pages}{002}
  (\bibinfo{year}{2003}), \eprint{astro-ph/0303636}.

\bibitem[{\citenamefont{{Gordon} and {Hu}}(2004)}]{GorHu04}
\bibinfo{author}{\bibfnamefont{C.}~\bibnamefont{{Gordon}}} \bibnamefont{and}
  \bibinfo{author}{\bibfnamefont{W.}~\bibnamefont{{Hu}}},
  \bibinfo{journal}{\prd} \textbf{\bibinfo{volume}{70}},
  \bibinfo{pages}{083003} (\bibinfo{year}{2004}), \eprint{astro-ph/0406496}.

\bibitem[{\citenamefont{{Bean} and {Dore}}(2004)}]{BeaDor04}
\bibinfo{author}{\bibfnamefont{R.}~\bibnamefont{{Bean}}} \bibnamefont{and}
  \bibinfo{author}{\bibfnamefont{O.}~\bibnamefont{{Dore}}},
  \bibinfo{journal}{\prd} \textbf{\bibinfo{volume}{69}},
  \bibinfo{pages}{083503} (\bibinfo{year}{2004}), \eprint{astro-ph/0307100}.

\end{thebibliography}

\end{document}